# Evaluation of Hash Algorithm Performance for Cryptocurrency Exchanges Based on Blockchain System

Abel C. H. Chen
*Information & Communications Security Laboratory,*
*Chunghwa Telecom Laboratories*
Taoyuan, Taiwan
ORCID: 0000-0003-3628-3033; Email: chchen.scholar@gmail.com

*Abstract*—The blockchain system has emerged as one of the focal points of research in recent years, particularly in applications and services such as cryptocurrencies and smart contracts. In this context, the hash value serves as a crucial element in linking blocks within the blockchain, ensuring the integrity of block contents. Therefore, hash algorithms represent a vital security technology for ensuring the integrity and security of blockchain systems. This study primarily focuses on analyzing the security and execution efficiency of mainstream hash algorithms in the Proof of Work (PoW) calculations within blockchain systems. It proposes an evaluation factor and conducts comparative experiments to evaluate each hash algorithm. The experimental results indicate that there are no significant differences in the security aspects among SHA-2, SHA-3, and BLAKE2. However, SHA-2 and BLAKE2 demonstrate shorter computation times, indicating higher efficiency in execution.

## I. INTRODUCTION

In recent years, blockchain systems [1] have matured, facilitating numerous applications and services, including cryptocurrencies [2] and smart contracts [3]. Among these, Bitcoin [4] and Ethereum [5] are widely recognized cryptocurrencies with significant commercial value. For instance, as of March 14, 2024, Bitcoin is trading at $73,449.10 USD per unit [6], while Ethereum is trading at $3,980.43 USD per unit [7], reflecting the market's enthusiasm for cryptocurrencies. However, with the advancement of quantum computing technology, there is a potential for disruption to existing cryptographic techniques [8]. Thus, the development of secure cryptographic technologies is crucial for maintaining and bolstering confidence in cryptocurrency markets.

In cryptocurrency exchanges based on blockchain systems, each block typically consists of a block hash, a previous block hash, transactions, generation time, and a nonce $n$ (shown in Fig. 1) [9]. The block hash and previous block hash play pivotal roles in constructing the blockchain. Additionally, hash values serve to verify the integrity of transactions and generation time. Moreover, while quantum computing may pose a threat to asymmetric cryptography, it currently does not present a viable means to compromise hash functions. Hence, employing hash functions remains a secure method for constructing blockchains. In efforts to enhance blockchain security, mechanisms such as Proof of Work (PoW) have been proposed. This involves setting a target value, denoted as $k$, and continually computing hash values until the first $k$ bits of the hash output are all zeros, at which point the nonce value $n$ is determined (shown in Fig. 2) [10]. For instance, let the block content be denoted as $c$, and define the hash algorithm as $Hash(x)$ with input $x$. The $i$-th hash value, $h_i$, is determined recursively by $Hash(h_{i-1})$. In this scenario, the hash value $h_n$ satisfies the condition that the first $k$ bits of the hash output are all zeros. Although the PoW mechanism can ensure integrity, the security of the PoW mechanism may be influenced by different hash algorithms.

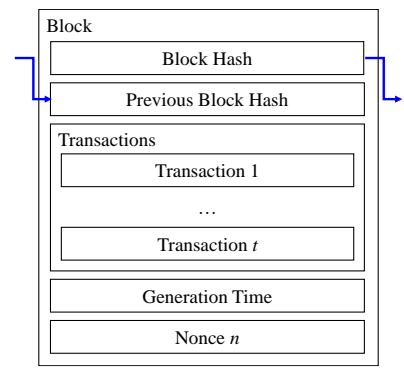

Fig. 1. A block in blockchain system.

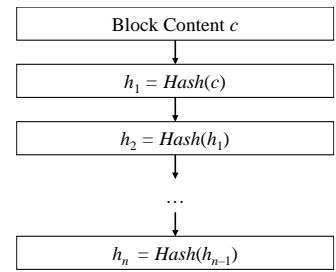

Fig. 2. The proof of work in blockchain system.

In light of this, the present study primarily investigates various hash algorithms (e.g. Secure Hash Algorithm (SHA)), comparing their respective security levels and performance within the context of PoW. The main contributions of this study are outlined as follows:

- This study proposes an evaluation factor for analyzing hash value heterogeneity, which can be utilized to select hash algorithms exhibiting high levels of heterogeneity.

- Various hash algorithms are explored in the context of PoW within blockchain systems in this study. It aims to validate which hash algorithm can accommodate larger nonce values, thereby selecting hash algorithms with higher security.

- This study compares the execution efficiency of different hash algorithms, analyzing the computational time required for each hash operation.
- The primary focus of this study is on comparing hash algorithms such as SHA-1, SHA-2, SHA-3, and BLAKE.

This paper comprises six sections. Section II elaborates on the concepts of hash algorithms. Section III introduces the evaluation factor and examines the hash value heterogeneity of each hash algorithm. Section IV analyzes the number of executions of each hash algorithm (i.e. the nonce value) for PoW within blockchain systems, while Section V presents the computation time for each hash algorithm. Finally, Section VI discusses the conclusions drawn from this study and outlines potential avenues for future research.

## II. SECURE HASH ALGORITHM

The flow of the hash algorithm, comprising (1) the initial setup and appending phase, (2) the confusion, substitution, and shift phase, and (3) the output phase, is depicted in Figure 3 [11]. This section provides a detailed exposition of the steps involved in SHA-1 in Subsection II.A, SHA-2 in Subsection II.B, and SHA-3 in Subsection III.C.

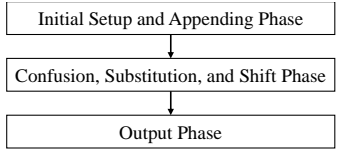

Fig. 3. The flow of the hash algorithm.

### A. SHA-1

In SHA-1 [12], the initial setup and appending phase involve the assignment of five 32-bit initial chain value, denoted as $C_0$. The message is padded to a multiple of 512 bits and segmented into $q$ 512-bit sub-messages. The $i$-th sub-message, denoted as $M_i$, is defined sequentially. During the confusion, substitution, and shift phase, a compression function $f_1(C_{i-1}, M_i)$ is established to compress the $i$-th sub-message through substitution and shift operations (as depicted in Eq. (1)). Each sub-message and chain value are inputted into the compression function iteratively until the $q$-th sub-message yields $C_q$, as illustrated in Eq. (2). Ultimately, the value of $C_q$ is output as the 160-bit hash value of the message during the output phase.

$$C_i = f_1(C_{i-1}, M_i) \tag{1}$$

$$C_q = f_1(C_{q-1}, M_q) \tag{2}$$

### B. SHA-2

SHA-2 comprises SHA-224 [13], SHA-256 [14], SHA-384 [15], and SHA-512 [16], with detailed descriptions of these hash algorithms provided in the subsequent paragraphs.

Taking SHA-256 as an example, its core framework is similar to SHA-1. However, in SHA-256, during the initial setup and appending phase, eight 32-bit initial chain value, denoted as $S_0$, are set to enhance security. Additionally, improvements are made to the compression function in SHA-256 to establish a more secure version, defined as $f_2(S_{i-1}, M_i)$, for compressing the $i$-th sub-message through substitution and shift operations (as depicted in Eq. (3)). Each sub-message and chain value are input into the SHA-256's compression function iteratively until the $q$-th sub-message yields $S_q$, as depicted in Eq. (4). Ultimately, the value of $S_q$ is output as the 256-bit hash value of the message during the output phase.

$$S_i = f_2(S_{i-1}, M_i) \tag{3}$$

$$S_q = f_2(S_{q-1}, M_q) \tag{4}$$

### C. SHA-3

SHA-3 was selected as a result of an open competition organized by the National Institute of Standards and Technology (NIST), with Keccak emerging as the chosen standard. Notably, the standard comprises SHA3-224 [17], SHA3-256 [18], SHA3-384 [19], and SHA3-512 [20]. Keccak primarily integrates sponge functions during the confusion, substitution, and shift phase, thereby enhancing the security of the compression function and allowing for flexible lengths of the output hash value.

Furthermore, BLAKE2 [21] was another algorithm considered during the selection process, demonstrating strong performance characteristics. As a result, this study incorporates BLAKE2 for comparative analysis.

## III. THE ANALYSIS OF HASH VALUE HETEROGENEITY

In this section, the evaluation factor is proposed in Subsection III.A, and two scenarios involving the immediately previous hash value and the m hash values for evaluation are provided separately in Subsections III.B and III.C, respectively.

### A. The Proposed Evaluation Factor

To analyze hash value heterogeneity, determining sufficient differences and confusion between each hash value, this section proposes a method of computing the XOR (exclusive-OR) result of two hash values. Subsequently, the number of bits set to 1 in this result is statistically calculated, representing the number of differing bits. The probabilities of differing and identical bits are then determined, followed by the utilization of entropy to measure randomness. If the probability of identical bits is 100%, it indicates complete similarity between the two hash values, resulting in an entropy of 0. Conversely, if the probability of differing bits is 100%, it implies complete dissimilarity, making the values susceptible to prediction by attackers, also yielding an entropy of 0. Ideally, the probability of differing bits should be 50%, with the probability of identical bits also at 50%, resulting in maximum entropy.

The specific procedure is as follows: Assuming the $i$-th hash value is $r_i$, the difference between $r_i$ and $r_j$, denoted as $d_{i,j}$, is calculated using Eq. (5) based on XOR operator $\oplus$. The number of bits set to 1, $b_1(d_{i,j})$, is then determined, and based on the hash value length $L$ and Eq. (6), the probability of differing bits, $p_{i,j,1}$, is computed. Finally, using Eq. (7), entropy $E_{i,j}$ is calculated to evaluate the quality of the hash algorithm. A higher entropy indicates better performance of the hash algorithm.

$$d_{i,j} = r_i \oplus r_j \tag{5}$$

$$p_{i,j,1} = b_1(d_{i,j}) / L \tag{6}$$

$$E_{i,j} = -[(1 - p_{i,j,1}) \times \ln(1 - p_{i,j,1}) + p_{i,j,1} \times \ln(p_{i,j,1})] \tag{7}$$

## B. Difference From the Immediately Previous Hash Value

Considering that PoW in blockchain involves iterative calculations of hash values, this section primarily explores the differences and entropy (i.e. $E_{i,i-1}$) between the $i$-th hash value and the $(i$-1)-th hash value. Subsequently, 32,768 hash values are generated for comparison and statistical analysis. This study mainly compares SHA-1, SHA-2 (including SHA-224, SHA-256, SHA-384, and SHA-512), SHA-3 (including SHA3-224, SHA3-256, SHA3-384, and SHA3-512), and BLAKE2 (including BLAKE2-256, BLAKE2-384, and BLAKE2-512). Experimental results are presented in Table I and Fig. 4. According to the experimental results, considering the median values, SHA-1 exhibits the poorest performance, while SHA-512, SHA3-512, and BLAKE2-512 demonstrate the best performance. Additionally, the performance of SHA-2, SHA-3, and BLAKE2 algorithms are almost identical.

TABLE I. DIFFERENCE FROM THE IMMEDIATELY PREVIOUS HASH VALUE BASED ON EACH HASH ALGORITHM

| Hash Algorithm | The 1st quartile | Median | The 3rd quartile |
|---|---|---|---|
| SHA-1 | 0.6928 | 0.6919 | 0.6893 |
| SHA-224 | 0.6930 | 0.6922 | 0.6899 |
| SHA-256 | 0.6929 | 0.6924 | 0.6907 |
| SHA-384 | 0.6930 | 0.6925 | 0.6915 |
| SHA-512 | 0.6930 | 0.6927 | 0.6919 |
| SHA3-224 | 0.6930 | 0.6922 | 0.6899 |
| SHA3-256 | 0.6929 | 0.6924 | 0.6907 |
| SHA3-384 | 0.6930 | 0.6925 | 0.6915 |
| SHA3-512 | 0.6930 | 0.6927 | 0.6919 |
| BLAKE2-256 | 0.6929 | 0.6924 | 0.6907 |
| BLAKE2-384 | 0.6930 | 0.6925 | 0.6915 |
| BLAKE2-512 | 0.6930 | 0.6927 | 0.6919 |

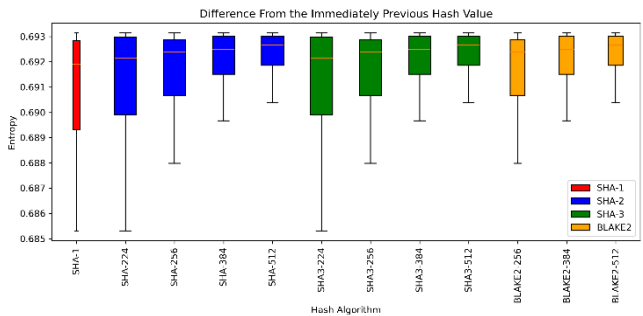

Fig. 4. Difference from the immediately previous hash value based on each hash algorithm.

For ease of comparison among the current mainstream algorithms, this section presents SHA-256, SHA3-256, and BLAKE2-256 in Fig. 5. From the experimental results, it can be observed that SHA-2, SHA-3, and BLAKE2 show no significant differences in performance.

## C. Difference From the m Hash Values

To compare the likelihood of hash collisions generated by different hash algorithms, this study conducts pairwise comparisons of the 32,768 generated hash values to compute the minimum entropy between each hash value and other hash values, as illustrated in Eq. (8). As mentioned earlier, this study primarily contrasts SHA-1, SHA-2 (including SHA-224, SHA-256, SHA-384, SHA-512), SHA-3 (including SHA3-224, SHA3-256, SHA3-384, SHA3-512), and BLAKE2 (including BLAKE2-256, BLAKE2-384, BLAKE2-512). The experimental results are presented in Table II and Fig. 6.

$$E_i^* = \min_{1 \le j \le 32768, i \ne j} E_{i,j} \qquad (8)$$

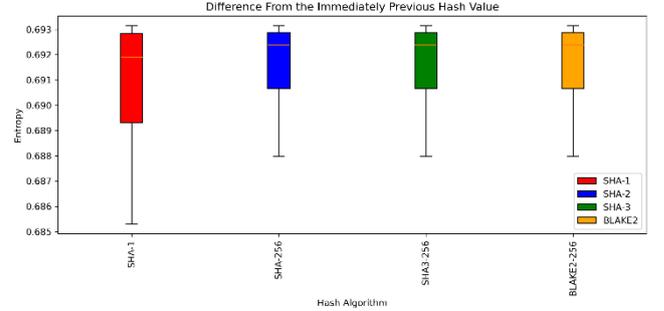

Fig. 5. Difference from the immediately previous hash value based on mainstream hash algorithms.

TABLE II. DIFFERENCE FROM THE $m$ HASH VALUES BASED ON EACH HASH ALGORITHM

| Hash Algorithm | The 1st quartile | Median | The 3rd quartile |
|---|---|---|---|
| SHA-1 | 0.6394 | 0.6351 | 0.6306 |
| SHA-224 | 0.6568 | 0.6518 | 0.6491 |
| SHA-256 | 0.6595 | 0.6574 | 0.6553 |
| SHA-384 | 0.6713 | 0.6690 | 0.6679 |
| SHA-512 | 0.6769 | 0.6755 | 0.6740 |
| SHA3-224 | 0.6568 | 0.6518 | 0.6491 |
| SHA3-256 | 0.6595 | 0.6574 | 0.6553 |
| SHA3-384 | 0.6713 | 0.6690 | 0.6679 |
| SHA3-512 | 0.6769 | 0.6755 | 0.6740 |
| BLAKE2-256 | 0.6595 | 0.6574 | 0.6553 |
| BLAKE2-384 | 0.6713 | 0.6690 | 0.6679 |
| BLAKE2-512 | 0.6769 | 0.6755 | 0.6740 |

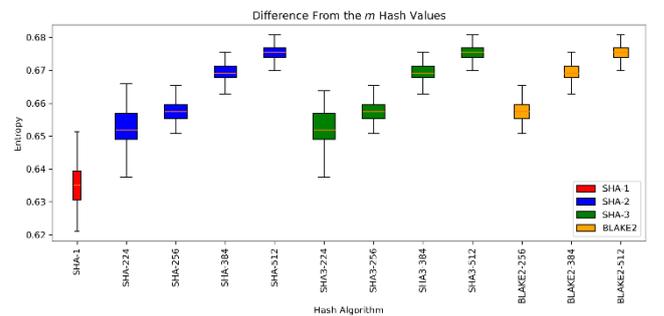

Fig. 6. Difference from the $m$ hash values based on each hash algorithm.

From the experimental findings, it can be observed that as a hash value is compared with more hash values to calculate its entropy, its minimum entropy will be lower than the entropy value calculated only with its adjacent two hash values. Furthermore, it is evident that as the length of the hash value increases, the entropy value also increases, indicating a reduced likelihood of collisions. Since the hash value length of SHA-1 is only 160 bits, it exhibits the poorest performance. In contrast, SHA-512, SHA3-512, and BLAKE2-512 have a length of 512 bits, resulting in the best performance. To compare mainstream hash algorithms, Fig. 7 illustrates the comparison results for SHA-256, SHA3-256, and BLAKE2-256. The experimental results indicate that there are no significant differences among SHA-2, SHA-3, and BLAKE2.

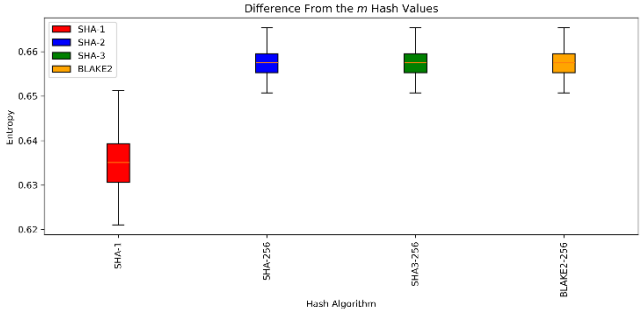

Fig. 7. Difference from the *m* hash values based on mainstream hash algorithms.

## IV. THE ANALYSIS OF POOF OF WORK

This section primarily analyzes the nonce value *n* used in PoW, where the nonce value primarily indicates the number of iterations of hash values required to satisfy the condition that the first *k* bits of the hash output are all zeros. If nonce values are too low, it implies that every device (including potential attackers) can easily complete the task, resulting in lower security. Therefore, nonce values should be distributed as uniformly as possible while being sufficiently large. In this section, the value of *k* is set to 8, and the performance of SHA-1, SHA-2 (including SHA-224, SHA-256, SHA-384, SHA-512), SHA-3 (including SHA3-224, SHA3-256, SHA3-384, SHA3-512), and BLAKE2 (including BLAKE2-256, BLAKE2-384, BLAKE2-512) is compared under the same conditions. The experimental results are presented in Fig. 8 and Table III. The results show no significant differences in the performance of each hash algorithm in this metric. However, BLAKE2-512 exhibits slightly higher values at the median. Therefore, when considering increasing the hash value to 512 bits in the future, adopting the BLAKE2 algorithm may be worth considering.

Fig. 9 selects mainstream algorithms, namely SHA-256, SHA3-256, and BLAKE2-256, and the experimental data indicate no significant differences among them.

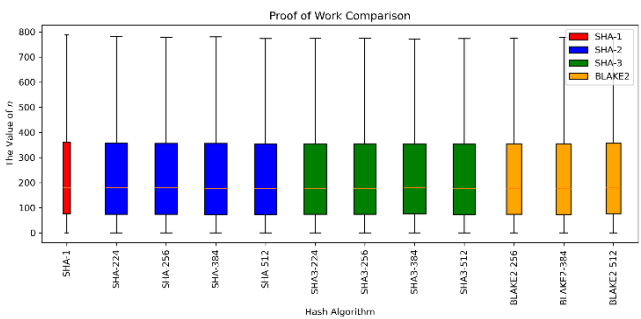

Fig. 8. The number of executions of each hash algorithm for the proof of work in blockchain system.

TABLE III. THE NUMBER OF EXECUTIONS OF EACH HASH ALGORITHM FOR THE PROOF OF WORK IN BLOCKCHAIN SYSTEM

| Hash Algorithm | The 1st quartile | Median | The 3rd quartile |
|---|---|---|---|
| SHA-1 | 361 | 179 | 76 |
| SHA-224 | 357 | 179 | 74 |
| SHA-256 | 356 | 179 | 74 |
| SHA-384 | 356 | 176 | 73 |
| SHA-512 | 353 | 178 | 73 |
| SHA3-224 | 354 | 178 | 74 |
| SHA3-256 | 355 | 178 | 74 |
| SHA3-384 | 354 | 179 | 75 |
| SHA3-512 | 353 | 177 | 73 |
| BLAKE2-256 | 355 | 177 | 74 |
| BLAKE2-384 | 355 | 177 | 73 |
| BLAKE2-512 | 357 | 180 | 76 |

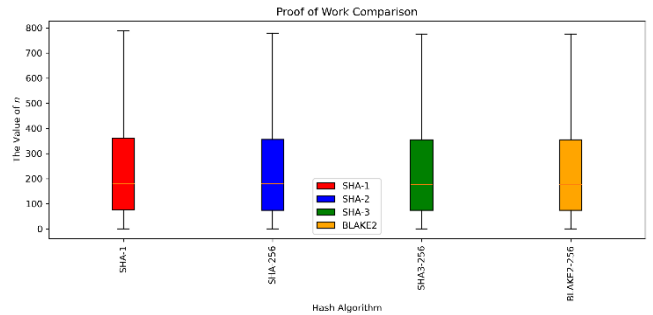

Fig. 9. The number of executions of mainstream hash algorithms for the proof of work in blockchain system.

## V. THE ANALYSIS OF COMPUTATION TIME

In the field of information security, security takes precedence over efficiency. Efficiency is considered secondary when equivalent security levels are achieved. Based on the experimental results from previous sections, it is evident that SHA-2, SHA-3, and BLAKE2 exhibit no significant differences in terms of security. Therefore, this section focuses on comparing their computational efficiency.

Table IV and Fig. 10 present the comparative experimental results of SHA-1, SHA-2 (including SHA-224, SHA-256, SHA-384, SHA-512), SHA-3 (including SHA3-224, SHA3-256, SHA3-384, SHA3-512), and BLAKE2 (including BLAKE2-256, BLAKE2-384, BLAKE2-512). The experimental results indicate that SHA-2 and BLAKE2 have shorter computation times, whereas SHA-3 requires longer computation times. Hence, in scenarios prioritizing efficiency, adopting SHA-2 and BLAKE2 may be preferable.

For the comparison of mainstream algorithms, Fig. 11 presents SHA-256, SHA3-256, and BLAKE2-256. From the experimental results, it is evident that SHA-256 and BLAKE2-256 exhibit higher efficiency compared to SHA3-256.

## VI. CONCLUSIONS AND FUTURE WORK

This study primarily focuses on evaluating the performance of various hash algorithms on the basis of a blockchain system. Additionally, it proposes measurement metrics to validate hash value heterogeneity. Based on the findings of this study, it is observed that SHA-2, SHA-3, and BLAKE2 exhibit no significant differences in terms of security. However, SHA-2 and BLAKE2 demonstrate higher execution efficiency. Therefore, when deploying applications involving cryptocurrencies in the future, consideration may be given to adopting SHA-2 and BLAKE2. Furthermore, according to the experimental results presented in Subsection III.C, increasing the hash value length effectively enhances security. Hence, it is recommended to utilize SHA-512 and BLAKE2-512 algorithms in the future.

TABLE IV. THE COMPUTATION TIME OF EACH HASH ALGORITHM (UNIT: NANOSECONDS)

| Hash Algorithm | *The 1st quartile* | *Median* | *The 3rd quartile* |
|---|---|---|---|
| SHA-1 | 1700 | 1100 | 300 |
| SHA-224 | 500 | 400 | 300 |
| SHA-256 | 500 | 300 | 300 |
| SHA-384 | 600 | 500 | 400 |
| SHA-512 | 1100 | 400 | 400 |
| SHA3-224 | 1400 | 1200 | 1100 |
| SHA3-256 | 2400 | 2100 | 1200 |
| SHA3-384 | 3700 | 2200 | 1300 |
| SHA3-512 | 3500 | 2100 | 1200 |
| BLAKE2-256 | 600 | 500 | 500 |
| BLAKE2-384 | 700 | 700 | 600 |
| BLAKE2-512 | 1500 | 1200 | 600 |

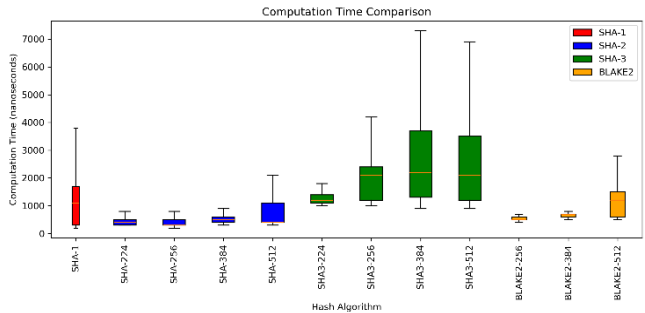

Fig. 10. The computation time of each hash algorithm (Unit: nanoseconds).

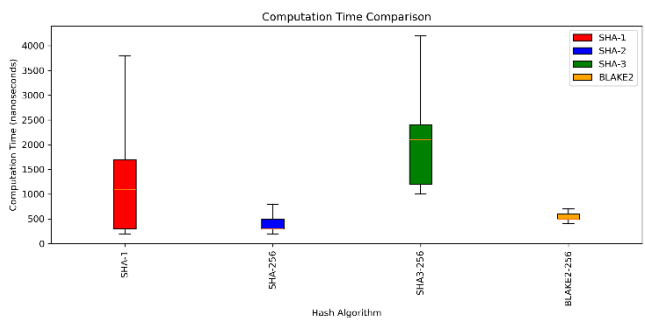

Fig. 11. The computation time of mainstream hash algorithms (Unit: nanoseconds).